\documentclass[10pt]{blois07}

\usepackage{graphicx}

\usepackage{cite,./mcite}

\newcommand{\be}{\begin{equation}}
\newcommand{\ee}{\end{equation}}
\newcommand{\bea}{\begin{eqnarray}}
\newcommand{\eea}{\end{eqnarray}}
\newcommand{\brho}{\mbf{\rho}}
\newcommand{\mbf}[1]{\mbox{\boldmath $#1$}}
\begin{document}
\title{Small $x$ QCD and Multigluon States: a Color Toy Model }

\author{G.P. Vacca$^1$\protect\footnote{\hspace{0.2cm}Talk presented at EDS07}
and P. L. Iafelice}
\institute{$^1$ INFN - Sez. di Bologna and Univ. of Bologna,
Physics Depart., Via Irnerio 46, 40126 Bologna
}
\maketitle
\begin{abstract}
We introduce and study a toy model with a finite number of degrees of freedom whose
Hamiltonian presents the same color structure of the BKP system appearing in the studies
of QCD in the Regge limit. We address within this toy model the question of
the importance of finite $N_c$ corrections with respect to the planar limit case. 
\end{abstract}
\section{Introduction}
The large $N_c$ expansion~\cite{tHooft} is a widely popular framework of
approximations which has been succesfully applied to gauge theories and has given at
leading order some analytical results otherwise impossible to obtain.
Within the Regge limit of QCD scattering amplitudes, L.N. Lipatov
found~\cite{integr} that systems of reggeized gluons evolving in rapidity
in the leading logarithmic approximation (LLA) were showing the emergence
of an integrable structure in the planar limit. Similar feature were found
later in other kinematical regimes for other QCD observables.
Moreover the $N=4$ SYM theory has been investigated at different order in
perturbation theory and is now believed to be integrable at all orders. 

But if one considers some QCD observables at the physical point $N_c=3$ the
situation is much more complicated and even the order of the corrections with
respect to the planar limit are not really known. This is the situation, for
example, for the spectrum of the BKP kernel~\cite{bkp1,bkp2}
at one loop, which describes the
high energy behavior in the Regge limit of a system of reggeized gluons.

It is the purpose of this talk to discuss a toy model~\cite{IV} which has a
color structure similar to the BKP system but a different ``configuration''
dynamics with a finite number of d.o.f., constrained only by the fact that the two
Hamiltonians must have the same leading eigenvalues in the large $N_c$ limit
for both one and two cylinder topologies.
The main motivation to study this model is to understand in a simpler case how much the
large $N_c$ approximation fails to reproduce the dynamics at finite $N_c$.
In order to understand this we shall study the spectrum of such model as a
function of $N_c$. 
\section{Small $x$ QCD: the LLA BKP kernel}
Let us start by giving a brief overview of the LLA kernels encoding the
evolution in rapidity of systems of interacting reggeized gluons, which
provide a convenient perturbative description of some relevant
QCD degrees of freedom in the Regge limit (small $x$). Their dynamics
determine the high energy behavior of the cross sections, typically
associated to the so called BFKL (perturbative) pomeron~\cite{bfkl1,bfkl2}. 
In the simplest form, the BFKL pomeron turns out to be a composite state of
two interacting reggeized gluons ``living'' in the transverse configuration plane
in the colorless configuration.
Its kernel or Hamiltonian is infrared finite and
in LLA is constructed summing the perturbative contributions of different
Feynman diagrams: in particular the virtual ones (reggeized one loop gluon trajectories)
$\omega$ and the real ones (associated to an effective real gluon emission
vertex) $V$. One
writes formally $H=\omega_1+\omega_2+ \vec{T}_1 \vec{T}_2 V_{12}$
where $\vec{T}_i$ are the generators of the color group in adjoint
representation . In the colorless case one has $\vec{T}_1 \vec{T}_2 =-N_c$
and finally one obtains:
\be
H_{12}=\ln \,\left| p_{1}\right| ^{2}+\ln \,\left| p_{2}\right| ^{2}+\frac{1%
}{p_{1}p_{2}^{\ast }}\ln \,\left| \rho _{12}\right| ^{2}\,p_{1}p_{2}^{\ast
}\,+\frac{1}{p_{1}^{\ast }p_{2}}\ln \,\left| \rho _{12}\right|
^{2}\,p_{1}^{\ast }p_{2}-4\Psi (1)\,,
\label{BFKL_mom}
\ee
where $\Psi (x)=d\ln \Gamma (x)/dx$, a factor $\bar{\alpha}_s =\alpha_s
N_c/\pi$ has been omitted and the gluon
holomorphic momenta and coordinates have been introduced.
One has the freedom, because of gauge invariance
to choose a description within the M\"obius space~\cite{lipatov-moebius,BLV2,BLSV}.
Then the BFKL hamiltonian has the property of the holomorphic
separability ($H_{12}=h_{12}+\bar{h}_{12}$) and is invariant under the
M\"obius group $SL(2,C)$ transformations, whose generators for the holomorphic sector
in the M\"obius space for the principal series of unitary representations
are given by
$M_{r}^{3}=\rho _{r}\partial _{r}\,,\,\,M_{r}^{+}=\partial
_{r}\,,\,\,M_{r}^{-}=-\rho^2_{r}\partial _{r}$.
The associated Casimir operator for two gluons is $
M^2 = |\vec{M}|^2=-\rho _{12}^2\,\partial _1\,\partial _2$
where $\vec{M}=\sum_{r=1}^{2}\vec{M}_{r}$ and $\vec{M}_r\equiv(M_r^+,M_r^-,M_r^3)$.
Note that, after defining formally $J(J-1)=M^2$, one may write
$h_{12}=\psi(J)+\psi(1-J)-2\psi(1)$.

The eigenstates and eigenvalues of the full hamiltonian in
eq. (\ref{BFKL_mom}), $H_{12} E_{h,\bar{h}} = 2 \chi_h E_{h,\bar{h}}$ are
labelled by the conformal weights
$h=\frac{1+n}{2}+i\nu$ ,\,\,$\bar{h}=\frac{1-n}{2}+i\nu$.
The leading eigenvalue, at the point $n=\nu=0$, has a value
$\chi_{max}=4\ln 2\approx 2.77259$, responsible for the rise of the total cross
section as $s^{\bar{\alpha}_s \chi_{max}}$, which corresponds to a strong
violation of unitarity.

Let us now consider the evolution in rapidity of composite states of
more than $2$ reggeized gluons~\cite{bkp1,bkp2}.
The BKP Hamiltonian in LLA, acting on a colorless state,
can be written in terms of the BFKL pomeron Hamiltonian and
has the form (see \cite{integr})
\be
H_n=-\frac{1}{N_c} \sum_{1\leq k < l\leq n} \vec{T}_{k} \vec{T}_{l}
H_{kl}\,.
\label{bkp_K}
\ee
This Hamiltonian is conformal invariant but can be solved only for $3$
reggeized gluons, since the color structure factorizes, leaving an
integrable dynamics~\cite{integr}. Different families of odderon solutions
were found~\cite{JW,BLV}. The family of solutions given in~\cite{BLV} are the
leading ones corresponding to intercept up to $1$ and have a non null coupling
to photon-meson impact factors~\cite{BBCV}.

The case of more than three reggeized gluon is in general not solvable but in
the large $N_c$ limit, taking the one cylinder topology (1CT), one obtains the
integrable Hamiltonian
\be
H_n^{\infty} = \frac{1}{2} \left[ H_{12}+ H_{23}
+ \cdots + H_{n1} \right] = h_n+\bar{h}_n \,,
\label{bkpnlarge}
\ee
i.e. there exists a set of other $n-1$ operators $q_r$,
which commute with it and are in involution.
This integrable model is a non compact generalization of the Heisenberg XXX
spin chains~\cite{integr,FK} and has been intensively studied with different
techniques in the last decade~\cite{dkm,dv-lip,dkkm,dv-lip2,vacca,kot}.

Here we remind the value of the highest eigenvalue of a system of $4$
reggeized gluons:
$H_4^{\infty} \psi_4=2 E_4^{\,1 \rm CT} \psi_4$.
The maximum value found, for zero conformal spin, is
\be
E_4^{\,1 \rm CT}=0.67416 \,.
\label{eigen1CT}
\ee
How to go beyond the large $N_c$ approximation is not an easy question to answer.
One may be tempted to apply variational or perturbative techniques to the
spectral problem, which nevertheless appears to be quite involved.
\section{Color structure for the $4$ gluon case}
\label{sec:cs4}
Let us analyze the color structure of the BKP kernel $H_4$ for four gluons, given in
eq. (\ref{bkp_K}). It is acting on 4-gluon states, which may be
represented as functions of the transverse plane coordinates and of the
gluon colors $v(\{\brho_i\})^{a_1 a_2 a_3 a_4}$.
Since the four gluons are in a total color singlet the color vector
$v^{a_1 a_2 a_3 a_4}$ can be described in terms of the color state of a two
gluon subchannel. On such a subspace, introducing the projectors
$P[R_i]_{a_1 a_2}^{a'_1 a'_2}$ onto irreducible representations of $SU(N_c)$, one has $
1=P_1+P_{8A}+P_{8S}+P_{10+\bar{10}}+P_{27}+P_{0} = \sum_i P[R_i]$,
where $Tr P[R_i]=d_i$ is the dimension of the corresponding representation and
we consider a unique subspace for the  $10$ and $\bar{10}$ representations. This
is convenient for our purposes and we shall therefore work with $6$
different projectors to span the color space of two gluons.  

On considering gluons $(1,2)$ to be the reference channel we introduce as the
base for the color vector space the set $\{ P[R_i]_{a_1 a_2}^{a_3 a_4}\}$ of
projectors and write
\be
v^{a_1 a_2 a_3 a_4}=\sum_i v^i \left(P[R_i]_{a_1 a_2}^{a_3 a_4}\right)  \quad
\mathrm{or} \quad v=\sum_i v^i P_{12}[R_i]\,.
\ee
Having chosen a color basis, the next step is to write the BKP kernel with
respect to it. Since $\sum_i \vec{T}_i v=0$ one may finally obtain:
\be
H_4=-\frac{1}{N_c} \left[ \vec{T}_1\vec{T}_2
  \left(H_{12}+H_{34}\right)+\vec{T}_1\vec{T}_3 \left(H_{13}+H_{24}\right)+
\vec{T}_1\vec{T}_4 \left(H_{14}+H_{23}\right) \right].
\label{bkp_K_2}
\ee
Let us now write explicitely the action of the color operators $\vec{T}_i\vec{T}_j=\sum_a
T^a_i T^a_j$ which are associated to the interaction between the gluons
labelled $i$ and $j$.
We start from the simple ``diagonal channel'' for which we have relation
$\vec{T}_i\vec{T}_j= -\sum_k a_k P_{ij}[R_k]$ with coefficients
$a_k=(N_c,\frac{N_c}{2},\frac{N_c}{2},0,-1,1)$.
Consequently we can write in the $(1,2)$ reference base
\be
\left(\vec{T}_1\vec{T}_2 v \right)^j=- a_j v^j = -\left( A \,v \right)^j \,,
\label{action12}
\ee
where $A=diag(a_k)$.
The action on $v$ of the $\vec{T}_1\vec{T}_3$ and $\vec{T}_1\vec{T}_4$
operators is less trivial and is constructed in terms of the $6j$ symbols of the
adjoint representation of $SU(N_c)$ group:
\be
\left( \vec{T}_1\vec{T}_3 \,v \right)^j=- \sum_i \left( \sum _k C^j_k a_k
    C^k_i\right) v^i = -\left( {C A\, C}\, v\right)^j
\label{action13}
\ee
and
\be
\left( \vec{T}_1\vec{T}_4 \,v \right)^j=- \sum_i \left( \sum _k s_j C^j_k a_k
    C^k_i s_i \right) v^i = -\left( {S C A\, C S}\, v\right)^j \,.
\label{action14}
\ee
The matrix $C$ is the symmetric crossing matrix build on the $6j$ symbols and
$S=diag(s_j )$ is constructed on the parities $s_j=\pm 1$ of the different
representations $R_j$.

We can therefore write the general BKP kernel for a four gluon state,
given in eq. (\ref{bkp_K_2}), as
\be
H_4=\frac{1}{N_c} \left[ A \left(H_{12}+H_{34}\right)+C A
  C\left(H_{13}+H_{24}\right)+
S C A C S \left(H_{14}+H_{23}\right) \right]
\label{H4colordecomp}
\ee
One can check that if we make trivial the transverse space dynamics, replacing
the $H_{ij}$ operators by a unit operators, the general BKP kernel in
eq. (\ref{bkp_K}) becomes $H_n= \frac{n}{2} \hat{1}$ and indeed one can verify
that $A+C A C+ S C A C S=N_c \hat{1}$.

Let us make few considerations on the large $N_c$ limit approximation. As we
have already discussed, in the Regge limit one faces the factorization of an
amplitude in impact factors and a Green's function which exponentiates the
kernel.
The topologies resulting from the large $N_c$ limit depend on the impact factor
structure. In particular one expects the realization of two cases: the one
and two cylinder topologies. The former corresponds to the case, well studied, of the
integrable kernel, Heisenberg XXX spin chain-like. It is encoded in the relation:
$\vec{T}_i \vec{T}_j\to -\frac{N_c}{2}\delta_{i+1,j}$ which leads to
 $H_4=\frac{1}{2}\left(H_{12}+H_{23}+H_{34}+H_{41}\right)$.
It is characterized by eigenvalues corresponding to an intercept less then a pomeron.
The latter case instead is expected to have a leading intercept, corresponding
to an energy dependence given by two pomeron exchange. Consequently one
expects at finite $N_c$ a contribution with an energy dependence even stronger. 
In the two cylinder topology the color structure is associated to two singlets
($\delta_{a_1a_2}\delta_{a_3a_4}$, together with
the other two possible permutations). Such a structure is indeed present in
the analysis, within the framework of extended generalized LLA, of unitarity
corrections to the BFKL pomeron exchange~\cite{Bartels} and diffractive
dissociation in DIS~\cite{BW}, where the perturbative triple pomeron
vertex (see also~\cite{BV,vaccaphd}) was discovered and shown to couple exactly to the
four gluon BKP kernel. 

It is therefore of great importance to understand how much the picture derived
in the planar $N_c=\infty$ case is far from the real situation with
$N_c=3$. One clearly expects for example that the first corrections to the
eigenvalues of the BKP kernel are proportional to $1/N_c^2$, but what is
unknown is the multiplicative coefficient as well as the higher order terms. 
\section{A BKP toy model}
In this section we shall consider a toy model~\cite{IV}, different from
the BKP system, but sharing several features with it and analysis within it
if the large $N_c$ approximation might be more or less satisfactory.

Besides the color space, a state of $n$ reggeized gluons undergoing the BKP evolution
belongs to the configuration space $R^{2n}$, associated to the position or
momenta in the transverse plane of the $n$ gluons.
The operators $H_{kl}$ act (see eq. (\ref{H4colordecomp})) on such a state and,
on the M\"obius space, can be written in terms of the
Casimir of the M\"obius group, i.e. in terms of the scalar product of the
generators of the non compact spin group $SL(2,{C})$: $H_{kl}=H_{kl}(
\vec{M}_{k}\cdot \vec{M}_{l})$. 

We are therefore led to consider a class of toy models where the BKP configuration
space $R^{2 n}$ is substituted by the space $V_s^n$ where $V_s$ is the
finite space spanned by spin states belonging to the irreducible
representation of $SU(2)$ with spin $s$. In particular we shall consider quantum
systems with an Hamiltonian:
\be
{\cal H}_n=-\frac{1}{N_c} \sum_{1\leq k < l\leq n} \vec{T}_{k} \vec{T}_{l}\,
f(\vec S_k \vec S_l)\,,
\label{bkp_toy_K}
\ee
where $\vec S_i$ are $SU(2)$ generators associated to the particle $i$ in
any chosen representation and $f$ is a generic function. A particular toy
model is therefore specified by giving the spin $s$ of each particle
(``gluons'') and the function $f$. Our BKP toy model is built choosing
the spin $s=1$ case in a global singlet state $v$ ($\sum_i \vec{S}_i \, v=0$)
and the family of functions $f$
\be
f_\alpha(x)=
2{\rm Re}\left[ \psi \left(\frac{1}{2}+\sqrt{-\alpha (4+2 x)}\right)\right]
-2\psi(1) \,.
\label{f_form}
\ee
This form is suggested by the conformal spin $n=0$ BFKL
Hamiltonian  with the substitution $\frac{1}{4}+L^2_{ij} \to - \alpha\,
S_{ij}^2$ which assures to have the same leading eigenvalue for any $\alpha$,
since both expressions have the value zero as upper bound. The parameter $\alpha$ will be
chosen in order to constrain the full 4-particle Hamiltonian (\ref{bkp_toy_K})
to have the same leading eigenvalue as the QCD BKP system in the large $N_c$
limit an one cylinder topology (at zero conformal spin), given in eq. (\ref{eigen1CT}).
This ``BKP toy model'' will be used to investigate finite $N_c$ effects. 

Since we have chosen to work with states singlet under $SU(2)_{spin\,\, conf}$
also for the spin part we employ the 2 particle subchannel decomposition in
irreducible representations, in a way similarly adopted for the color part.
After that one is left with the problem of diagonalizing an Hamiltonian which
is a matrix $18 \times 18$. Therefore we proceed by introducing for 2 particle
spin $1$ states the resolution of unity $1=Q_1+Q_3+Q_5=\sum_i Q[R_i]$ which let us write
$f(\vec{S}_i \vec{S}_j)=\sum_k f(-b_k) Q_{ij}[R_k]$ with $b_k=(2,1,-1)$. 
using a power series representation ($Q_{ij}[R_k]$ are
projectors). Introducing the crossing matrices $D$ and the parity matrix $S'$
we obtain the relations
$\left(f\left(\vec{S}_1\vec{S}_2 \right)v \right)^j=\left( B \,v \right)^j$,
$ \left( f\left(\vec{T}_1\vec{T}_3\right) \,v \right)^j= \left( {D B\, D}\,v\right)^j$
and 
$\left(f\left( \vec{T}_1\vec{T}_4 \right)\,v \right)^j=
\left( {S' D B\, D S'}\, v\right)^j$.
It is then straightforward to derive a matrix form for the Hamiltonian of this
toy model
\be
{\cal H}_{4a}=\frac{2}{N_c}\left( A\otimes B+C A\, C \otimes D B\, D+   S C A\, C S
\otimes  S' D B\, D S' \right)
\label{H4BKPtoy}
\ee
which depends on $N_c$ and on the parameter $\alpha$ through the
function $f_\alpha$ given in eq. (\ref{f_form}).

In the large $N_c$ limit one faces for the Hamiltonian two
possible cases (see~\cite{IV} for more details):
the one cylinder topology (1CT) which corresponds to the
simpler Hamiltonian ${\cal H}_{4a}^{1CT}= B+S'DBDS'$ and the two
cylinder topology (2CT) corresponding to the even simpler Hamiltonian
${\cal H}_{4a}^{2CT}=2B$.
Let us remark that while in the case of $N_c>3$ we consider a basis for the
vector states made of
{$P[R_i]Q[R_j]$} with $18$ elements since in the color sector there is also
the $P_0$ projector, the case $N_c=3$ is characterized by a basis of $15$
elements.

The last step to obtain the BKP toy model is to fix the parameter $\alpha$ by
requiring ${\cal H}_{4a}^{1CT}$ to have the value of eq. (\ref{eigen1CT}) so
we obtain $\alpha=2.80665$.
We are therefore left with an Hamiltonian which is just a function of the
number of colors $N_c$. 

Let us now consider its spectrum for the cases $N_c=3$
and $N_c=\infty$. Here we report just the leading eigenvalues of
${\cal H}_{4a}$ with their multiplicities:
($7.042$, $2\times 5.519$, $2\times 1.123$, $\cdots$). 
Changing $N_c$ from $3$ to $\infty$ we observe that the first three move to
the 2CT leading eigenvalue $5.545$ while the next two move
to the 1CT leading eigenvalue $0.674$.
With very good approximation one finds that the $N_c$ dependence of the
leading eigenvalue $E_0$ is given by
\be
E_0(N_c)=E_0(\infty)\left(1+\frac{2.465}{N_c^2}\right) \,.
\ee
One can see that for this toy model the large $N_c$
approximation corresponds to an error of about $27\%$, an error which is not negligible
because the coefficient of the leading correction to the asymptotic value, proportional to
$1/N_c^2$, is a large number.
The color- ``spin'' configuration mixing which is encoded in the eigenvectors has
been also studied. 
\section{Conclusions}
We have introduced a family of dynamical models describing interacting particles with
color and spin degrees of freedom in order to see how much
the large $N_c$ approximation is significant when one is trying to
extract the spectrum of these quantum systems.
In particular we have investigated a toy model, constructed to mimic some
features of the $4$ gluon BKP system. We have determined the $N_c$ dependence
of the spectrum and discussed the $N_c=\infty$ limit finding for the
leading eigenvalue corrections of about $30\%$ at $N_c=3$.

%
%
%
\begin{footnotesize}

\bibliographystyle{blois07} 

{\raggedright

\bibliography{blois07}

}

\end{footnotesize}

\end{document}